\documentstyle[times,pramana,epsf,floats]{ias}
\def\lsim{\mathrel{\rlap{\lower4pt\hbox{\hskip1pt$\sim$}}
    \raise1pt\hbox{$<$}}}                
\def\gsim{\mathrel{\rlap{\lower4pt\hbox{\hskip1pt$\sim$}}
    \raise1pt\hbox{$>$}}}                
\begin{document}
\mark{{Transport properties ...}{R.S. Bhalerao}}
\title{Transport properties of the fluid produced at RHIC}

\author{Rajeev S. Bhalerao}
\address{Department of Theoretical Physics, Tata Institute of Fundamental 
Research, Homi Bhabha Road, Colaba, Mumbai 400005}
\keywords{Relativistic heavy-ion collisions, quark-gluon plasma, transport 
properties}
\pacs{25.75.-q, 12.38.Mh}
\abstract{
It is by now well known that the relativistic heavy-ion collisions at
RHIC, BNL have produced a strongly interacting fluid with remarkable
properties, among them the lowest ever observed ratio of the
coefficient of shear viscosity to entropy density. Arguments based on
ideas from the String Theory, in particular the AdS/CFT
correspondence, led to the conjecture --- now known to be violated ---
that there is an absolute lower limit $1/4 \pi$ on the value of this
ratio. Causal viscous hydrodynamics calculations together with the
RHIC data have put an upper limit on this ratio, a small multiple of
$1/4 \pi$, in the relevant temperature regime. Less well-determined is
the ratio of the coefficient of bulk viscosity to entropy
density. These transport coefficients have also been studied
nonperturbatively in the lattice QCD framework, and perturbatively in
the limit of high-temperature QCD. Another interesting transport
coefficient is the coefficient of diffusion which is also being
studied in this context. I review some of these recent developments
and then discuss the opportunities presented by the anticipated LHC
data, for the general nuclear physics audience.}

\maketitle
\section{Introduction}

Transport coefficients of the QCD matter are of fundamental importance
not only because they represent an important aspect of QCD, but also
because they can be calculated from first principles. Trying to
extract these coefficients reliably from experimental data and
evaluating them in various theoretical approaches is a very active
area of research today.

This review is addressed to the {\it general Nuclear Physics} audience,
a majority of whom {\it do not work} in the area of relativistic
heavy-ion collisions, but have probably heard the claim that the RHIC
experiments have produced {\it the most perfect fluid} ever observed.
Before I explain the meaning of this claim and present the
experimental evidence for it, a few introductory remarks are in order.
It may be recalled that the phase diagram (pressure vs temperature) of
the most familiar liquid, namely water is known for long with good
accuracy. In particular, the coordinates of the triple point and the
critical point are known to several significant places,
and the various phase co-existence lines are well-determined. In
contrast, the phase diagram of the strongly-interacting matter or the
QCD phase diagram (temperature vs the net baryon number density or
equivalently the baryon number chemical potential) is known only
schematically from the experimentalist's point of view. Indeed, it is
not even known with certainty what are the various phases that occur
at high densities.

The big idea is to map out quantitatively the QCD phase diagram with
the relativistic heavy-ion collisions as an experimental tool. Such
experiments have been performed at SPS (CERN), are being performed at
RHIC (BNL), and will soon be performed at LHC (CERN), at successively
higher energies: up to $\sqrt{s}_{NN}=$ 19, 200 and 5400 GeV, at the
above three facilities, respectively. It is also necessary to
systematically scan the energy range up to 200 GeV in order to study
the QCD matter at high baryon number density and to locate the
critical point and the phase transition line predicted by some
theories. This is being done at RHIC and plans are afoot to do it at
FAIR (GSI) and NICA (JINR).

In nonrelativistic fluid dynamics, the kinematic viscosity ($\nu$) is
defined as $\nu=\eta/\rho$ where $\eta$ is the coefficient of shear
viscosity or the dynamic viscosity and $\rho$ is the density of the
fluid. It allows us to compare the viscosities of fluids with
different densities. (Interestingly, under standard conditions, water
has a {\it lower} $\nu$ than air, although its $\eta$ is higher.) The
dimensionless ratio $\eta/s$ serves as the relativistic analog of the
kinematic viscosity, $s$ being the entropy density.\footnote{Strictly
speaking, relativistic analog of $\rho$ is $(\epsilon+P)=sT$ where
$\epsilon$ is the energy density and $P$ the pressure.} (The
dimensionless ratio $\eta/n$ where $n$ is the number density is of no
use here because $n$ is ill-defined in relativistic heavy-ion
collisions.)

\begin{figure}[htbp]
\epsfxsize=6cm
\centerline{\epsfbox{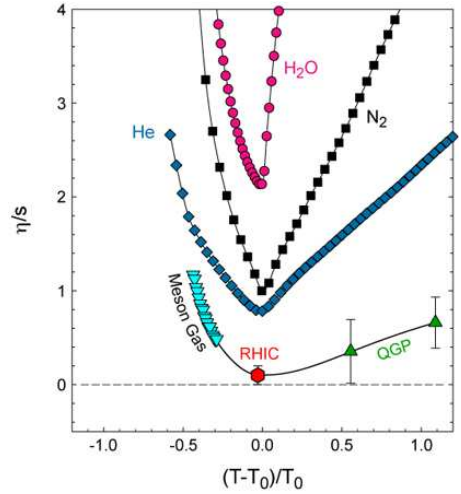}}
\smallskip
\caption{Comparison of different fluids with each other.
See the text for details. The lines are drawn to guide the eye.
Figure from Ref. \protect\cite{Lacey:2006bc}.}
\label{fig:lacey}
\end{figure}
Figure \ref{fig:lacey} shows $\eta/s$ as a function of $(T-T_0)/T_0$
where $T$ is the temperature and $T_0$ the critical temperature, for
various fluids, namely water, nitrogen, helium and the QCD matter. The
upper three curves, drawn at the respective critical pressures,
exhibit a cusp at $T_0$. The liquid and gaseous phases behave
differently because the momentum transport mechanisms are different in
the two cases; see, e.g., \cite{Csernai:2006zz}. The points labelled
{\it Meson Gas} have large ($\sim 50$ \%) errors and are obtained from
chiral perturbation theory. The points labelled {\it QGP} are from
lattice QCD simulations of \cite{Nakamura:2004sy}. The point labelled
{\it RHIC} is discussed below. This comparison of various fluids
explains the statement that the RHIC fluid is the most perfect fluid
ever observed.

\begin{figure}[htbp]
\epsfxsize=4cm
\centerline{\epsfbox{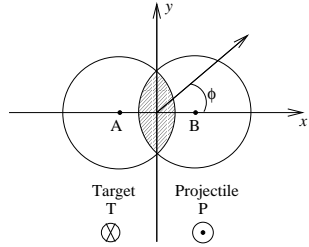}}
\caption{Non-central collision of two nuclei.}
\label{fig:colli}
\end{figure}

The point labelled {\it RHIC} in Fig. \ref {fig:lacey} was obtained by
matching the elliptic flow data at RHIC with the results of viscous
hydrodynamic calculations: Consider a non-central collision of two
identical spherical nuclei as shown in Fig. \ref{fig:colli}. The
nuclei travel parallel to the $z$ axis, $xy$ plane is the azimuthal or
transverse plane and $xz$ plane is called the reaction plane. The
overlap zone is shown as the shaded area in the figure. In an
ultrarelativistic collision, the nucleons in the non-overlapping zones
continue to travel more or less along their pre-collision
trajectories, leaving behind the almond-shaped overlap zone. The
interesting observables are governed mostly by the overlap zone which
has a very high initial energy density. The {\it spatial anisotropy}
of the overlap zone ensures anisotropic pressure gradients in the $xy$
plane. This leads to a final state characterized by {\it momentum
anisotropy} and anisotropic distribution of particles in the $xy$
plane.

The triple differential invariant distribution of particles emitted in
the final state is given by
\[
E\frac{d^3N}{d^3p}=\frac{d^3N}{p_Tdp_Tdyd\phi}
=\frac{d^2N}{p_Tdp_Tdy}\frac{1}{2\pi} \left[ 1+ \sum_1^\infty
2 v_n \cos(n\phi) \right],
\]
where $p_T$ is the transverse momentum, $y$ the rapidity and $\phi$
the azimuthal angle of an emitted particle. The azimuthal distribution
is Fourier-decomposed, and the leading coefficients $v_1$ and $v_2$
are called the directed and elliptic flow, respectively. They provide
a measure of the anisotropy of the flow in the transverse plane,
mentioned above. The importance of $v_2$ lies in the fact that it is a
measure of pressure at early times and hence the measure of
thermalization of the quark-gluon matter produced in heavy-ion
collisions.

\begin{figure}[htbp]
\epsfxsize=10cm
\centerline{\epsfbox{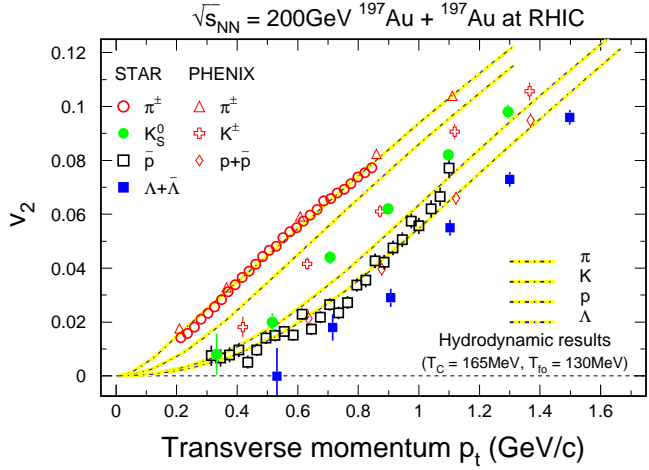}}
\caption{Success of ideal hydrodynamics: Minimum-bias elliptic flow
data compared with ideal hydrodynamics calculations of
Huovinen. Figure from Ref. \protect\cite{Oldenburg:2004qa}.}
\label{fig:v2}
\end{figure}

Hydrodynamics is an effective theory that describes the slow,
long-wavelength motion of a fluid close to equilibrium. It is a
powerful technique because given the initial conditions and only the
equation of state (EoS) of the matter, it predicts the space-time
evolution of the fluid. Its limitation is that it is applicable at or
near (local) thermodynamic equilibrium only. Hydrodynamics plays a
central role in modeling relativistic heavy-ion collisions.
State-of-the-art calculations for the RHIC data are based on the
relativistic causal dissipative hydrodynamics. However, the initial
calculations were done in the ideal ($\eta/s=0$) hydrodynamics
framework. As an example, see Fig. \ref {fig:v2}. The broad agreement
between the data and these initial calculations, in particular the
mass ordering of $v_2(p_T)$, led to the claim of formation of an ideal
fluid at RHIC.

\section{Transport Coefficients}

Consider a fluid in equilibrium. If it is perturbed so that the
density is no longer uniform, it responds by setting up currents which
tend to restore the equilibrium. In the linear response theory, the
current or flux is proportional to the force which is negative of the
gradient of the density: $\rho {\bf u}=-D \nabla \rho$. The constant of
proportionality ($D$) is called the coefficient of diffusion. Other
familiar examples are Ohm's law ${\bf J}=\sigma {\bf E}=-\sigma \nabla
\phi$, Fourier's law of heat conduction ${\bf Q}=-\kappa \nabla T$,
etc. A slightly more complicated example involves the transport of
momentum in response to velocity gradients in an anisotropic medium,
the constant of proportionality being the shear viscosity tensor. Such
equations are called constitutive equations because they express
physical properties of the material concerned. They relate the fluxes
with the forces, the constants of proportionality being the transport
coefficients.

In addition to the variables such as the hydrodynamic four-velocity,
pressure, energy density, conserved-number density, etc.,
hydrodynamics equations also contain transport coefficients (shear and
bulk viscosities, thermal conductivity, relaxation times, etc.). These
are external parameters which can be calculated in a variety of ways
(see Table 1) and fed into the hydrodynamics equations.

\begin{table}[h]
\caption{Various ways to calculate transport coefficients for 
relativistic fluids}
\begin{center}
\begin{minipage}{4.4in}
\begin{tabular}{|l|l|l|} \hline\hline
Weak-coupling regime & Kinetic Theory & Boltzmann Equation\\ \cline{2-3}
        & Linear-Response Theory & Kubo Formula \\ \cline{2-3}
              & N=4 Supersymmetric Yang-Mills Theory & \\ \hline
Strong-coupling regime & Lattice Gauge Theory & Kubo Formula \\ \cline{2-3}
  & N=4 Supersymmetric Yang-Mills Theory &  \\ \hline
\end{tabular}
\end{minipage}
\end{center}
\end{table}

\subsection{Transport coefficients from high-temperature QCD}

High-temperature QCD assumes $T \gg \Lambda_{QCD}$.
This is a weak-coupling regime and the shear ($\eta$) and bulk
($\zeta$) viscosities can be calculated in the kinetic theory
\cite{Arnold:2000dr},\cite{Arnold:2003zc},\cite{Arnold:2006fz}:
\[
\eta \sim \frac{T^3}{\alpha_s^2 \ln \alpha_s^{-1}},
~~~ \zeta \sim \frac{\alpha_s^2 T^3}{\ln \alpha_s^{-1}}.
\]
It is clear that as the temperature $T$ rises, $\eta/T^3$ increases
while $\zeta/T^3$ decreases, and the ratio $\zeta/\eta \sim
\alpha_s^4$ decreases. Note that the bulk viscosity vanishes for any
conformal field theory, and QCD becomes conformal in the limit of high
$T$. On the other hand, when $T \sim 200$ MeV, QCD is far from weakly
coupled and the above results can provide only a rough estimate of
$\eta$ and $\zeta$.

\subsection{Transport coefficients from string theory}

Like ordinary fluids black holes too are thermal systems having notions of
temperature and entropy. An object falling on the surface of a fluid
in equilibrium generates disturbance which dies down due to
dissipative nature of the fluid. Similarly, the black hole horizon
gets deformed when an object falls on it. However, it soon recovers
its equilibrium shape. Thus the notion of ``viscosity'' is applicable
to a black hole as well, and the connection between hydrodynamics and
black-hole physics does not seem very far-fetched.

Anti-deSitter/conformal field theory (AdS/CFT) correspondence refers
to the equivalence or duality between string theory defined on a
certain AdS space and a CFT defined on its boundary. It allows the
calculation of properties of a strongly-coupled CFT in terms of those
of a weakly-coupled string theory. For an elementary introduction to
these ideas, see \cite{Natsuume:2007qq}.

Ordinary quantum mechanics rules out vanishing $\eta/s$ for weakly
coupled theories, i.e., for theories with well-defined quasiparticles
\cite{Danielewicz:1984ww}. Kovtun et al. \cite{Kovtun:2004de}, using
string theory methods, showed that $\eta/s = 1 / 4 \pi$, for a large
class of strongly interacting quantum field theories whose dual
description involves black holes in the Anti-deSitter (AdS) space. The
value $1/4 \pi$ was conjectured to be its absolute lower bound (KSS
bound) for all substances. However, it has recently been realized that
the KSS bound is violated for certain conformal field theories (CFT)
\cite{Buchel:2008vz},\cite{Brigante:2008gz}.

Of course, applying above ideas to the fluid produced at RHIC is
speculative: QCD near the deconfinement transition temperature is not
a CFT, and its gravity dual (if it exists) is not known.

\subsection{Transport coefficients from lattice QCD}

\begin{figure}[htbp]
\epsfxsize=7cm
\centerline{\epsfbox{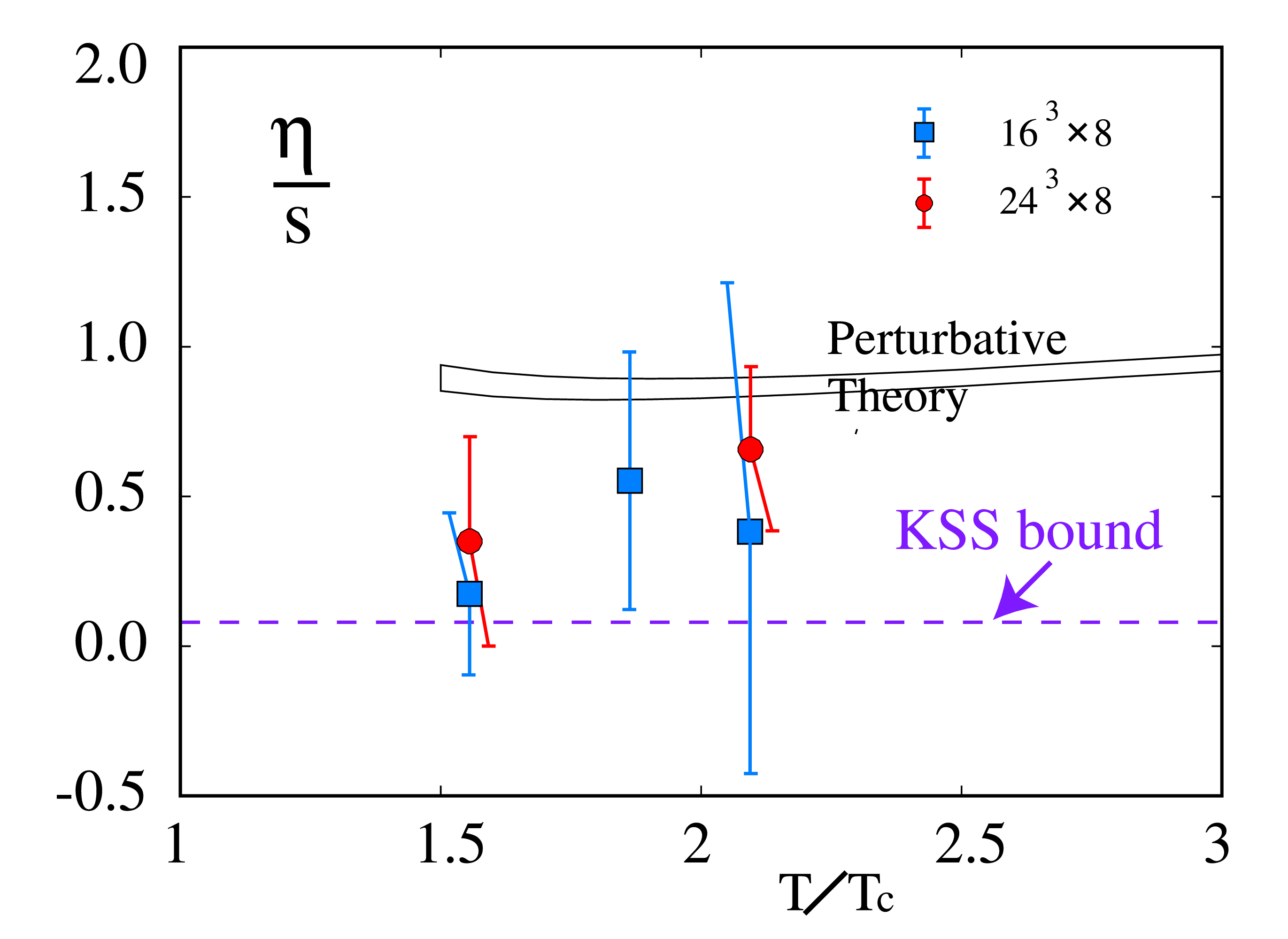}}
\caption{First lattice QCD results on $\eta/s$. Figure from Ref.
\protect\cite{Nakamura:2004sy}.}
\label{fig:lattice1}
\end{figure}

Figure \ref{fig:lattice1} shows the results of a quenched QCD (no
dynamical quarks or $m_q \rightarrow \infty$) calculation of $\eta/s$
from Ref. \cite{Nakamura:2004sy} in comparison with the high-$T$ QCD
results quoted in subsection 2.1 and the KSS bound mentioned in the
previous subsection. However, these initial results have now been 
superseded by more recent results described next. 

\begin{figure}[htbp]
\epsfxsize=7cm
\centerline{\epsfbox{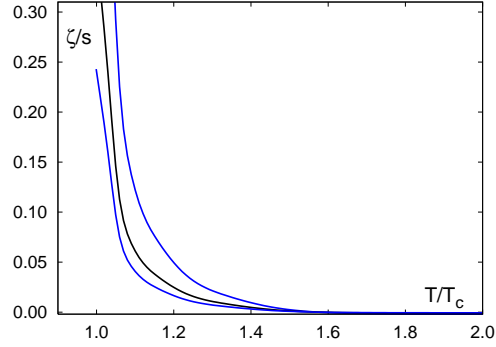}}
\caption{Bulk viscosity vs temperature. The three curves (top to
bottom) correspond to $\omega_0(T)=0.5,1,1.5$ GeV respectively, where
$\omega_0(T)$ is a scale at which the perturbation theory becomes
valid. The three curves reflect the uncertainty in the determination
of this scale parameter. Figure from
Ref. \protect\cite{Karsch:2007jc}. See, however, the discussion in the
text.}
\label{fig:lattice2}
\end{figure}

Recently Meyer has calculated both $\eta/s$ and $\zeta/s$ on the
lattice, with higher statistical accuracy and a more efficient
algorithm, assuming SU(3) gluodynamics
\cite{Meyer:2007ic},\cite{Meyer:2007dy}. He gets
\[
\eta/s = \left\{ 
\begin{array}{l@{~~~}l}
   0.134(33) & (T=1.65T_c) \\
   0.102(56) & (T=1.24T_c),
\end{array} \right.
\nonumber
\]
where the errors contain an estimate of the systematic uncertainty.
This is consistent with the KSS bound. Further,
\[
\zeta/s = \left\{ 
\begin{array}{l@{~~~}l}
  0.008(7)\Big[\begin{array}{c}\scriptstyle{0.15}\\ \scriptstyle{0} 
\end{array}\
\Big]
& (T=1.65T_c
) \nonumber  \\
  0.065(17)\Big[\begin{array}{c}\scriptstyle{0.37}\\ \scriptstyle{0.01}
\end{array}\Big]
 & (T=1.24T_c
),
           \end{array} \right.  \nonumber
\]
\[
\zeta/s = 0.73(3) 
\Big[\begin{array}{c}\scriptstyle{2.0}\\
\scriptstyle{0.5}\end{array}\Big]
\qquad   (T=1.02T_c
),  \nonumber
\]
where the statistical error is given and the square bracket specifies
conservative upper and lower bounds. Note the sharp rise in $\zeta/s$
just above $T_c$. Similar dramatic rise was seen in the results
presented in \cite{Karsch:2007jc}; see Figure \ref{fig:lattice2}.
They extracted the bulk viscosity in the presence of light quarks by
combining low-energy theorems with lattice data on the QCD EoS.
However, it is now realized that a determination of $\zeta$ from
correlation functions of the energy-momentum tensor is more subtle.
Shortcomings of this calculation have been pointed out in
\cite{Huebner:2008as}. If the sharp rise of $\zeta/s$ just above $T_c$
is confirmed, it would imply that the QGP is not a perfect fluid near
$T_c$!

For a review of the progress made in extracting transport properties
of the gluonic plasma from lattice simulations, see
\cite{Meyer:2009jp}.

\subsection{Transport coefficients extracted from RHIC data}

Here the basic idea is that the shear viscosity reduces the elliptic
flow: $v_2({\rm viscous~ fluid}) < v_2({\rm ideal~ fluid})$. This is
easy to understand: recall that $v_2$ is a measure of the flow
anisotropy in the azimuthal plane. Viscosity is the result of a
frictional force. Frictional force being proportional to the flow
velocity has a relatively stronger effect on fast-moving particles
emerging in the reaction plane. This reduces the anisotropy and hence
$v_2$.

\begin{figure}[htbp]
\epsfxsize=13cm
\centerline{\epsfbox{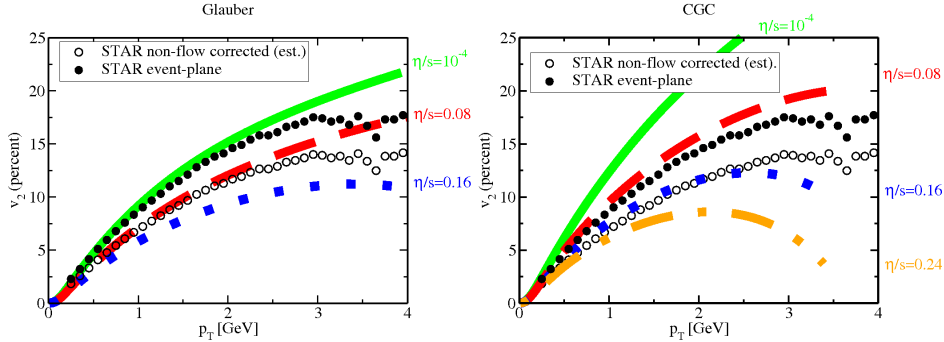}}
\caption{Minimum-bias elliptic flow data for charged hadrons in 200
GeV Au-Au collisions; only statistical errors are shown. Line
thickness is an estimate of the accumulated numerical error. Figure
from Ref. \protect\cite{Luzum:2008cw}.}
\label{fig:luzum}
\end{figure}

Thus if one has a good control on $v_2({\rm ideal~ fluid})$, one can
adjust $\eta/s$ to fit the data on $v_2$, and thus extract
$\eta/s$. This has been done by several groups \cite{Fries:2008ts},
\cite{Romatschke:2007mq}, \cite{Luzum:2008cw},
\cite{Song:2007ux},\cite{Dusling:2007gi},\cite{Molnar:2008xj},
\cite{Chaudhuri:2009uk}. Results of \cite{Luzum:2008cw} are shown in
Fig. \ref{fig:luzum}. The Glauber and colour-glass-condensate (CGC)
initial conditions for hydrodynamic evolution yield $\eta/s \simeq
0.08$ and $\simeq 0.16$ respectively, showing the sensitivity of the
extracted $\eta/s$ to the initial conditions.

\section{Diffusion}

Transport of heat occurs by diffusion (Fourier's law) as well as by
propagating waves (Maxwell-Cattaneo law). The latter mechanism
underlies the well-known phenomenon of second sound in superfluid
helium (Table 2). Similarly the transport of momentum too occurs by
diffusion (Navier-Stokes equation) or by propagating waves. These are
the transverse or shear waves\footnote{They are used, e.g., in
Magnetic Resonance Elastography (MRE) for quantitatively imaging
material properties.}, different from the more familiar longitudinal
or compressional waves. For a detailed discussion see
\cite{Romatschke:2009im}.

\begin{table}[h]
\caption{Various transport phenomena}
\begin{center}
\begin{minipage}{4.5in}
\begin{tabular}{|l|l|l|l|} \hline\hline
Transport of &  By diffusion & By propagating waves & 
Experimental situation\\
 \hline\hline
Heat & Fourier's law & Maxwell-Cattaneo law & Second sound 
in superfluid He \\ \hline
Momentum & Navier-Stokes eq.   & Maxwell-Cattaneo law & Propagating 
shear waves \\
\hline
Conserved no. & Fick's law & Kelly's law    & RHIC? \\
e.g., B, Q, S &      &          & \\ \hline
\end{tabular}
\end{minipage}
\end{center}
\end{table}

\begin{figure}[htbp]
  \epsfxsize=12.2cm
  \centerline{\epsfbox{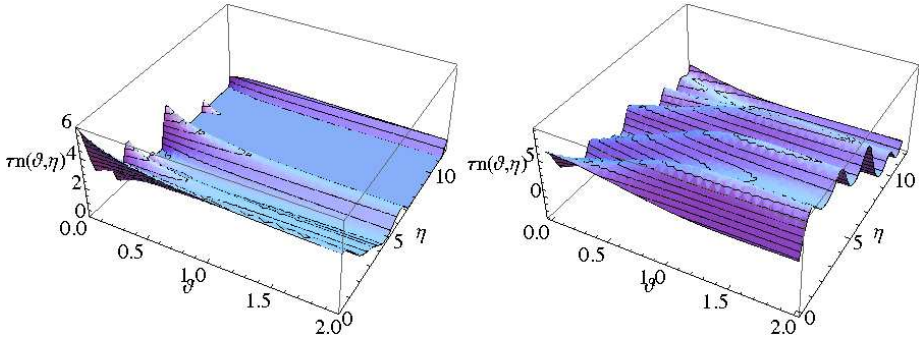}}
\medskip
  \caption{Conserved number density profile as a function of time and
space-time rapidity in the Fick (left) and Kelly (right) theories
of diffusion. Figure from Ref. \protect\cite{Bhalerao:2009tf}.}
  \label{fig:fick}
\end{figure}

We recently studied the transport of a conserved number such as the
net baryon number ($B$), charge ($Q$), or strangeness ($S$) conserved
in strong interactions, in the acausal and causal, or the first- and
second-order theories of relativistic diffusion
\cite{Bhalerao:2009tf}. We found that Fick's diffusion smooths out
gradients in the number density monotonically. In contrast, in Kelly's
theory the gradients may be transiently amplified, i.e., the density
profile may stiffen at intermediate times; see Fig. \ref {fig:fick}.
We proposed experimental observables and argued that the RHIC data can
potentially distinguish between the above two mechanisms. {\it If the
second-order theory of diffusion is ruled out by the data, one gets a
handle on the relaxation time and hence on thermalization.}

\section{Viscosity in Ordinary Finite Nuclei}

This review is about the transport properties of the fluid produced at
RHIC. But it is pertinent to ask what do we know about the transport
coefficients of the matter in an ordinary finite nucleus at low
temperatures. Hydrodynamic models have a long history in Nuclear
Physics --- recall the success of the liquid drop model. The
coefficient of shear viscosity of a finite nucleus can be obtained
from (i) analysis of the widths of giant resonances within the
hydrodynamic model, (ii) the process of fission studied within the
liquid drop model, and (iii) kinetic theory. This has been discussed
recently in \cite{Auerbach:2009ba}. Using entropy density for a
free Fermi gas or for noninteracting nucleons in a Woods-Saxon
potential, they obtained values of $\eta/s$ larger than but not
drastically different from those for the RHIC fluid.

\section {What about LHC?}

It is clear from Table 3 that at LHC we would need $\eta/s$ and
$\zeta/s$ up to $T \sim 4 T_c$. Some preliminary lattice results are
now available, but for the SU(3) pure gauge theory
(i.e., quarkless QCD)
\cite{Meyer:2009jp}. For a careful extraction of $\eta/s$ and
$\zeta/s$ from LHC (and RHIC) data, we need to incorporate the
$T$-dependence of these transport coefficients in hydrodynamic
calculations, among other refinements of these calculations; see the
next section.

\begin{table}[h]
\caption{Comparison of central Au-Au collisions at RHIC and central
Pb-Pb collisions at LHC}
\begin{center}
\begin{minipage}{2.8in}
\begin{tabular}{|l|l|l|} \hline\hline
 & RHIC (Au-Au) &       LHC (Pb-Pb)  \\   \hline\hline
$\sqrt{s}_{NN}$        & 200 GeV & 5.5 TeV  \\
Initial temperature    &   $\sim$2T$_{c}$ & $\sim$4T$_{c}$  \\
Initial energy density     & $\sim 5$ GeV/fm$^{3}$ &  15-60 GeV/fm$^{3}$  \\
Lifetime & $\sim 10$ fm/c  & $>$ 10 fm/c \\ \hline
\end{tabular}
\end{minipage}
\end{center}
\end{table}

\section{Take-Home Message}

\begin{itemize}

\item Elliptic flow at RHIC has put a robust {\it upper} limit on the
value of $\eta/s$ of the RHIC fluid: $\eta/s \lsim 5/(4\pi)$. This is the
average value in the relevant temperature region.

\item Uncertainties associated with the initial conditions, EoS, bulk
viscosity, hadronic stage, freezeout procedure, different versions of
the second-order hydrodynamic equations prevent a more precise
determination of $\eta/s$. See Ref. \cite{Heinz:2009cv} for a detailed
discussion.

\item The bulk viscosity is not yet well-determined in the
deconfinement transition region.

\item Analysis of the RHIC data, which might throw light on the
coefficient of diffusion, is awaited.

\end{itemize}

\bigskip\bigskip
\begin{center}
{\bf Acknowledgments}
\end{center}
I thank Sourendu Gupta and Jean-Yves Ollitrault for a critical reading
of the manuscript.


\end{document}